\documentclass[11pt]{article}
\usepackage[margin=.8in,left=.8in]{geometry}
\usepackage{amsmath}
\usepackage{amsfonts}
\usepackage{amssymb}
\usepackage{amsthm}
\usepackage{subcaption}
\usepackage{float}
\usepackage{graphicx}
\usepackage{color}
\usepackage{xcolor}
\usepackage[utf8]{inputenc}
\usepackage{hyperref}
\usepackage[all]{xy}

\usepackage{tikz}
\usepackage{tikz-cd}
\usepackage{lipsum}
\usepackage{adjustbox}

\usepackage{multirow}

\usepackage{stmaryrd}    % for \sslash

\usepackage{enumerate} %Roman enumerate

\usepackage{helvet}   %vertical spacing in tabular

\usepackage{mathptmx}
\usepackage{amsmath}
\usepackage{graphicx}

\usepackage{floatflt}  %Floating tables
\usepackage{wrapfig} %floating figure
\usepackage{array}     % specifying size of tables
\newcolumntype{L}[1]{>{\raggedright\let\newline\\\arraybackslash\hspace{0pt}}m{#1}}
\newcolumntype{C}[1]{>{\centering\let\newline\\\arraybackslash\hspace{0pt}}m{#1}}
\newcolumntype{R}[1]{>{\raggedleft\let\newline\\\arraybackslash\hspace{0pt}}m{#1}}

\usepackage{diagbox}  %diagonal in table

\makeatletter
\newcommand{\raisemath}[1]{\mathpalette{\raisem@th{#1}}}
\newcommand{\raisem@th}[3]{\raisebox{#1}{$#2#3$}}
\makeatother

\usepackage{tabularx}   %For sizing tabular entries

\usepackage[new]{old-arrows}   %For \longhookrightarrow

\newdir{> }{{}*!/10pt/@{>}}

\makeatletter
\newif\if@sup
\newtoks\@sups
\def\append@sup#1{\edef\act{\noexpand\@sups={\the\@sups #1}}\act}%
\def\reset@sup{\@supfalse\@sups={}}%
\def\mk@scripts#1#2{\if #2/ \if@sup ^{\the\@sups}\fi \else%
  \ifx #1_ \if@sup ^{\the\@sups}\reset@sup \fi {}_{#2}%
  \else \append@sup#2 \@suptrue \fi%
  \expandafter\mk@scripts\fi}
\def\tensor#1#2{\reset@sup#1\mk@scripts#2_/}
\def\multiscripts#1#2#3{\reset@sup{}\mk@scripts#1_/#2%
  \reset@sup\mk@scripts#3_/}
\makeatother

\makeatletter
\newbox\slashbox \setbox\slashbox=\hbox{$/$}
\def\itex@pslash#1{\setbox\@tempboxa=\hbox{$#1$}
  \@tempdima=0.5\wd\slashbox \advance\@tempdima 0.5\wd\@tempboxa
  \copy\slashbox \kern-\@tempdima \box\@tempboxa}
\def\slash{\protect\itex@pslash}
\makeatother

\def\clap#1{\hbox to 0pt{\hss#1\hss}}

\let\oldroot\root
\def\root#1#2{\oldroot #1 \of{#2}}
\renewcommand{\sqrt}[2][]{\oldroot #1 \of{#2}}

\DeclareSymbolFont{symbolsC}{U}{txsyc}{m}{n}
\SetSymbolFont{symbolsC}{bold}{U}{txsyc}{bx}{n}
\DeclareFontSubstitution{U}{txsyc}{m}{n}

\DeclareSymbolFont{stmry}{U}{stmry}{m}{n}
\SetSymbolFont{stmry}{bold}{U}{stmry}{b}{n}

\DeclareFontFamily{OMX}{MnSymbolE}{}
\DeclareSymbolFont{mnomx}{OMX}{MnSymbolE}{m}{n}
\SetSymbolFont{mnomx}{bold}{OMX}{MnSymbolE}{b}{n}
\DeclareFontShape{OMX}{MnSymbolE}{m}{n}{
    <-6>  MnSymbolE5
   <6-7>  MnSymbolE6
   <7-8>  MnSymbolE7
   <8-9>  MnSymbolE8
   <9-10> MnSymbolE9
  <10-12> MnSymbolE10
  <12->   MnSymbolE12}{}

\makeatletter
\def\Decl@Mn@Delim#1#2#3#4{%
  \if\relax\noexpand#1%
    \let#1\undefined
  \fi
  \DeclareMathDelimiter{#1}{#2}{#3}{#4}{#3}{#4}}
\def\Decl@Mn@Open#1#2#3{\Decl@Mn@Delim{#1}{\mathopen}{#2}{#3}}
\def\Decl@Mn@Close#1#2#3{\Decl@Mn@Delim{#1}{\mathclose}{#2}{#3}}
\Decl@Mn@Open{\llangle}{mnomx}{'164}
\Decl@Mn@Close{\rrangle}{mnomx}{'171}
\Decl@Mn@Open{\lmoustache}{mnomx}{'245}
\Decl@Mn@Close{\rmoustache}{mnomx}{'244}
\makeatother

\makeatletter
\DeclareRobustCommand\widecheck[1]{{\mathpalette\@widecheck{#1}}}
\def\@widecheck#1#2{%
    \setbox\z@\hbox{\m@th$#1#2$}%
    \setbox\tw@\hbox{\m@th$#1%
       \widehat{%
          \vrule\@width\z@\@height\ht\z@
          \vrule\@height\z@\@width\wd\z@}$}%
    \dp\tw@-\ht\z@
    \@tempdima\ht\z@ \advance\@tempdima2\ht\tw@ \divide\@tempdima\thr@@
    \setbox\tw@\hbox{%
       \raise\@tempdima\hbox{\scalebox{1}[-1]{\lower\@tempdima\box
\tw@}}}%
    {\ooalign{\box\tw@ \cr \box\z@}}}
\makeatother

\makeatletter
\def\udots{\mathinner{\mkern2mu\raise\p@\hbox{.}
\mkern2mu\raise4\p@\hbox{.}\mkern1mu
\raise7\p@\vbox{\kern7\p@\hbox{.}}\mkern1mu}}
\makeatother

\newcommand{\R}{\ensuremath{\mathbb R}}
\newcommand{\Z}{\ensuremath{\mathbb Z}}

\renewcommand{\(}{\begin{equation}}
\renewcommand{\)}{\end{equation}}
\newcommand{\bea}{\begin{eqnarray*}}
\newcommand{\eea}{\end{eqnarray*}}

\usepackage{cleveref}

\crefformat{section}{\S#2#1#3} % see manual of cleveref, section 8.2.1
\crefformat{subsection}{\S#2#1#3}
\crefformat{subsubsection}{\S#2#1#3}

\theoremstyle{italics}
\newtheorem{theorem}{Theorem}[section]

\newtheorem{prop}[theorem]{Proposition}

\theoremstyle{definition}

\newtheorem{remark}[theorem]{Remark}
\newtheorem{note[theorem]}{Note}

\usepackage{amsfonts}

\author{Hisham Sati}
\title{
Six-dimensional gauge theories 
and (twisted) generalized cohomology}  

\begin{document}
\maketitle 

\begin{abstract}

We consider the global aspects of the 6-dimensional $\mathcal{N}=(1, 0)$ 
theory arising from  the coupling of the vector multiplet to the tensor multiplet. 
We show that the Yang-Mills field and its dual, when both are abelianized, combine 
to define a class in twisted cohomology with the twist arising from the class of the 
$B$-field, in a duality-symmetric manner.  
We then show that this lifts naturally to a class in \emph{twisted} (differential) K-theory.
Alternatively, viewing the B-field in both $\mathcal{N}=(1,0)$ and $\mathcal{N}=(2,0)$ 
theories, not as a twist but as an invertible element, leads to a description within 
\emph{untwisted} chromatic level two generalized cohomology theories, 
including forms of elliptic cohomology and Morava K-theory. 
 \end{abstract}

\tableofcontents

%%%%%%%%%%%%%%%
\section{Introduction}
%%%%%%%%%%%%%%%%

\medskip
String theory constructions have given rise to many novel 
$6$-dimensional ${\cal N}=(1, 0)$ super-conformal field theories (SCFTs)
 \cite{Wi}\cite{St}\cite{SWi}\cite{HZ}. These  theories admit the following
bosonic field multiplets \cite{Se}: 
\begin{enumerate} 
\vspace{-2mm} 
\item \emph{Vector multiplet}:  
%The bosonic field content is only 
a vector field $A$. 
 \vspace{-2mm} 
\item  \emph{Hypermultiplet}: 
%The bosonic field content is 
4 real scalars.  
\vspace{-2mm} 
\item  \emph{Tensor multiplet}: 
%The bosonic field content is 
a two-form $B$ and a real scalar
$\varphi$. The field strength $H$ corresponding to $B$ is constrained to be self dual, 
that is $H = *H$.  
\end{enumerate} 
\vspace{-2mm} 
Such theories come in many varieties, and various physical approaches to 
their (physics) classifications have been  proposed \cite{HMV}\cite{Bh}\cite{HMRV}. 
These come equipped with additional global symmetries in comparison with
${\cal N}=(2, 0)$ theories.  Topological considerations have proven to be very rich 
and lead to connections to invariants of smooth 4-manifolds \cite{GPPV}.

\medskip
Approaching the $\mathcal{N}=(1,0)$ and $\mathcal{N}=(2,0)$ theories from more traditional 
physics angles have led to interesting descriptions (see \cite{Moore} and references therein). 
However, in order to uncover structures that exist behind such theories, one would want to  
pursue a study from a geometric and topological points of view. 
Generally, six-dimensional theories have led to interesting geometric and 
topological structures, especially ones on the 
M5-brane \cite{tcu} and on the NS5-brane \cite{NS5}. One of the main points 
we advocate here in this note is to change the viewpoint on the vector multiplet 
Yang-Mills fields, leading to novel structures and connections to twisted generalized 
cohomology theories. 
%, as we
%did for the heterotic fields in \cite{}. 

 \medskip
 There are some similarities between the $d=6$, ${\cal N}=(1,0)$ theory and the heterotic theory in  
 $d=10$, in the sense of existence of Yang-Mills fields with corresponding 
 Green-Schwarz anomaly cancellation phenomena due to coupling to other 
 fields. Indeed, one has the Green-Schwarz-West-Sagnotti anomaly cancellation in six-dimensional 
supergravity coupled to vector multiplets 
 \cite{GSW}\cite{GN}\cite{Sag}\cite{Ho}\cite{KMW}\cite{OSTY}\cite{In} \cite{MMP}.

\medskip
The tensor multiplet field $H_3$ interacts with the above Yang-Mills in an interesting fashion.
In our set-up, $H_3$ will take on the role of a twist for the above fields. 
However, the Yang-Mills fields being nonabelian are not 
directly amenable to a description using (twisted) generalized cohomology. Nevertheless, we adopt
the strategy from \cite{higher} to abelianize the fields which look otherwise like Yang-Mills fields
to bring out their generalized cohomology aspect. What makes the fields change nature in a sense
is the fact that they couple to other fields in a process akin to the Chapline-Manton coupling \cite{CM}.  
This is similar to we did for the heterotic case in \cite{higher}, 
and these abelian fields have recently been derived from first principles via super homotopy theory 
\cite{FSS9}. The proposal in \cite{higher} then led to 
the construction of twisted versions of generalized cohomology theories, namely Morava
K-theory and E-theory with higher twists \cite{SWe} and is an ingredient in the consistency
of the recently uncovered novel T-duality in M-theory \cite{SS}.

\medskip
Six-dimensional ${\cal N}=(1,0)$ supergravity theories 
with abelian  gauge group have been studied in \cite{PT}.   Such a context can also lead to constraints. 
For instance, in \cite{LRW}, it is shown that the structure 
of anomaly cancellation implies that an abelian gauge theory in six dimensions cannot exist in 
absence of gravity, in which case it is inconsistent as a gauge theory.  

%\medskip
%In analogy with the case of M5-branes (see \cite{tcu}\cite{Moore}), we expect the theory 
%to be rich from a geometric  and topological point of view.
%

\medskip
The abelianized gauge fields  
will be described at the level of differential forms, by the de Rham complex. We then promote 
to a description in de Rham cohomology and integral cohomology, upon invoking integrality conditions
traced back to the fact that their nature as gauge fields. These abelian Yang-Mills (or Maxwell) fields
will then be candidates for lifting to generalized cohomology theories, the most transparent being 
K-theory. We do so by demonstrating the vanishing of the corresponding obstructions.

\medskip
 While generally for the ${\cal N}=(2, 0)$
we have the class $H_3$, which would be a candidate for a twist, in that case there are no fields in sight 
to be twisted by it. In contrast, the ${\cal N}=(1, 0)$ provides a candidate for what would be twisted, 
namely the Yang-Mills fields arising from the vector multiplet. 

\medskip
From a conceptual perspective, one wonders whether there is a canonical description of the B-field 
on the worldvolume. From one angle, the work \cite{MS} looked at the effect of the B-field on the 
worldvolume as leading to noncommutativity resulting in a new perspective on S-duality. 
Here we provide another angle. We hope that by looking at different perspectives and assembling 
them, a more complete and coherent picture will ultimately emerge. 

\medskip
Alternatively, we could ask whether $H_3$ can be interpreted not as a a twist, but as 
an element in some untwisted cohomology theory. This was considered in the context of 
S-duality in type IIB string theory in \cite{KS2}, where twisted K-theory was demonstrated 
to not be compatible with S-duality and that an interpretation that goes beyond such 
a description is needed. One proposal there was to view it as an invertible element in some 
chromatic level two generalized cohomology theory. We find that similar arguments can 
be invoked in our current context of 6-dimensional theories.

\medskip
Since we are considering only one element, rather than a collection assembled via periodicity 
into a bigger total field of a uniform degree, the most natural approach is to view $H_3$ 
as an invertible element, i.e., as that special element given as a higher analogue of a line 
bundle. Being in cohomological degree three, the natural candidate theories will be at 
chromatic level two (as opposed to it being one for K-theory at chromatic level one). 
Such theories include Morava K-theory  $K(2)$ and forms of elliptic cohomology, 
such as Morava E-theory $E(2)$, algebraic K-theory of  the topological K-theory 
spectrum $K_{\rm alg}(Ku)$, and topological modular forms tmf. This part does not 
essentially involve the vector multiplet (except for cancelling mixed anomalies)
and so applies to both $\mathcal{N}=(1,0)$ and $\mathcal{N}=(2,0)$ theories. 

\medskip
Asking for a family of generalized cohomology theories to be twisted by a gerbe of 
any cohomological degree  has led to construction of twists of iterated algebraic 
K-theory of topological complex K-theory, which models higher vector 
bundles \cite{LSW}.

\medskip
This is also, indirectly, related to twisted spectra, as we will consider a map which 
is essentially a looping of the map defining the universal twist for the spectra under
consideration. 

\medskip
This note is organized as follows. In \cref{Sec-cases} we provide the setting in terms 
of abelianized fields, written as a components of a total field to be twisted, and action functionals, 
whose variations lead to equations of motion that we arrange in a suggestive manner 
to lend themselves to a geometric and topological interpretations in the following section. 
Indeed, in \cref{Sec-K} we interpret these expressions as the vanishing of certain twisted
differentials in twisted de Rham cohomology. We then lift these to twisted K-theory and then to 
twisted differential K-theory, demonstrating explicitly the vanishing of the obstructions,
which leads to state our main result. Finally, in \cref{Sec-un} we provide a second take on
the description of the fields in the tensor multiplet as invertible elements in untwisted 
chromatic level two cohomology theories. We also also highlight the relation to twisted 
K-theory from which we transition smoothly.

%%%%%%%%%%%%%%
\section{The abelian gauge theory and the combined field strength} 
\label{Sec-cases}
%%%%%%%%%%%%%%

 Considering the gauge field $F_2$ and its dual $F_4:=*F_2$ in the $\mathcal{N}=(0, 1)$ 
 theory as a unified field $\mathcal{F}$, we show that the equations of motion at the 
 rational level takes on a form that leads to a twisted differential.

\medskip
There is a similarity to  the heterotic theory in the existence of a link  between the $H$-field $H_3$
and the Chern-Simons form of the gauge theory via the Manton-Chapline coupling 
\cite{CM}\cite{BddN}.  In that setting, in \cite{higher} it was proposed that the gauge field $F_2$ 
and its 10-dimensional dual $F_8:=*_{10}F_2$, being related to $H_3$ that way, should lead 
to an interpretation  in terms of generalized cohomology.

\medskip
We can choose not to impose 
self-duality on $H_3$ or, alternatively, consider a pseudo-action
which yields the  equations of motions for the tensor fields which still need to be
additionally restricted by imposing  (anti-)selfduality constraints 
\cite{NS}\cite{FRS}\cite{RS}\cite{Ric}\cite{GSS}\cite{SSW1}. 
The action functional with which we start is (see \cite{KKP}) 
\(
S_6=\tfrac{1}{2}\int_{M^6}  H_3 \wedge * H_3 + \sqrt{c} \int_{M^6} 
\Big( -\phi {\rm tr}(F_2 \wedge *F_2) 
+ B_2 \wedge {\rm tr}(F_2 \wedge F_2) \Big) 
\)
where $*$ is the Hodge operator in six dimensions, $\phi$ is a scalar, 
and $c$ is a constant that depends on the gauge group and the number of flavors $N_f$. 
For ${\rm SU}(N)$, this is $c=4N - N_f$. The field $H_3$ is given by 
$H_3=dB_2 + \sqrt{c} CS_3(A)$, where $CS_3(A)$ is the Chern-Simons form 
for the gauge potential $A$ with the Yang-Mills field strength $F_2$.

\medskip
As in \cite{higher}, we would like to consider the case where the Yang-Mills group,  
$G$ in the classifications \cite{HMV} \cite{Bh}\cite{HMRV}\cite{MMP}\cite{MM}, is 
broken down to an abelian subgroup. That is, we are in the Coloumb branch of the 
6-dimensional theory.  We assume that the manifold $M^{6}$ is such 
that this breaking via Wilson lines is possible, say down to (a subgroup of)
 the Cartan torus. For the abelian case, the 
second Chern class is trivial in cohomology, so that the second Chern form ${\rm  tr}(F_2 \wedge F_2)$ 
becomes essentially $F_2 \wedge F_2$, as the trace results in valued-ness in an abelian lattice, but also that this 
4-form is exact and trivialized by a 3-form. Furthermore, we have $N=1$, so that $c=4-N_f$. This already 
restricts the number of flavors allowed.

\medskip
Let us then consider the resulting pseudo-action of the form
\footnote{We often drop numerical 
factors which do not affect the results we are after.}
\(
S=\int_{M^6} H_3 \wedge * H_3 + \int_{M^6} F_2 \wedge * F_2 + 
\int_{M^6} B_2 \wedge  F_2 \wedge F_2\;,
\label{originalaction} 
\)
with a Chapline-Manton coupling $H_3=CS_3(A)$, and where we have 
set the scalar field to have a constant value $\phi=-\tfrac{1}{2\;\sqrt{c}}$.
Of course, we could carry this with us with obvious multiplicative 
shifts in the fields below. 

\medskip
In order to work our way through, 
we consider two stages, depending on whether or not we include 
the anomaly term, which is either the last term in \eqref{originalaction} 
or the more general term $\int_{M^6} B_2 \wedge Y_4$, where $Y_4$ is a 
more general 4-form discussed below. Hence we allow for three cases.

\paragraph{Case 1: No anomaly term.} Varying the action \eqref{originalaction} 
(excluding the last term) with respect to $B_2$, we get the equation of motion 
\(
d*H_3=0\;.
\)
In addition, we have the `Bianchi identity' $dH_3=0$, also obtained from self-duality. 
Note that these equations are what one would also obtain in the $\mathcal{N}=(2,0)$ theory. 
Let us next vary the action with respect to  the gauge field $A$
 in order to get the gauge field equation of motion. 
 We have (assuming the abelian case)
\(
\frac{\delta S}{\delta A} = * H_3 \wedge F_2 - d* F_2 \;,
\)
which implies the equation for the gauge field,
\(
(d* - * H_3 \wedge) F_2=0\; .
\)
We would like to manipulate this equation to put it in 
a more suggestive form. Namely, in order to write the operator
as $(d - \mathcal{O})$, we define the combined or total curvature
\(
{\mathcal F}:=F_2 + * F_2=F_2 + F_4 
\)
of the gauge field strength and its dual. The gauge field equation
 will then be written as
\(
(d - * H_3 \wedge) {\mathcal F}=0\;.
\label{eq-dstarH} 
\)
This suggests that the combined object  ${\mathcal F}$ is the analog of the situation
in the case of the Ramond-Ramond (RR) fields $\mathcal{F}_{RR}$ in type II string theories
which combines all fields of the same $\Z_2$ parity (see \cite{Fr}\cite{tcu}), as well as 
of the heterotic gauge field and its dual  as $\mathcal{F}_{\rm het}=F_2 + F_8$ \cite{higher}.

\paragraph{Case 2: Inclusion of the gauge anomaly term.} Next, we add the anomaly 
term $\int_{M^6} B_2 \wedge F_2  \wedge F_2$ back into the action \eqref{originalaction}. 
In this case, the EOMs for the B-field and its dual are 
\(
\label{EOM-Bianchi}
d*H_3= c \;{\rm tr} (F_2 \wedge F_2) 
\qquad 
\text{and} 
\qquad 
dH_3= c \; {\rm tr} (F_2\wedge F_2)\;.
\)
As indicated above, we need to abelianize the anomaly term ${\rm tr}(F_2 \wedge F_2)$. 
Again, this is the second Chern form, which should be exact, reflecting the fact that the corresponding 
class $c_2$ is trivial in cohomology. 

\medskip
While this is somewhat restrictive, we can nevertheless arrange for $F_2\wedge F_2$ 
to be zero as a differential form. 
For instance, in the limiting case of flat space $\R^6$ with coordinates $x, y, z, u, v, w$, 
the 2-form $F_2=dx \wedge dy$ wedge-squares to zero. More generally, we can take
$F_2$ to be reduced in the following sense. Take $F_2 \in \Lambda^2 \R^6$, but 
consider a vector subspace $\R^m \subset \R^6$, $m < 6$, such that 
$F_2 \wedge F_2 \in \Lambda^m \R$.  If we make $m$ to be
smaller than the rank of the form $F_2 \wedge F_2$, i.e. $m < 4$, then $F_2 \wedge F_2=0$ 
as a form. We can transparently extend this to nontrivial manifolds, where now 
$F_2 \in \Lambda^2 TM$ for a manifold $M$. While we can arrange for $F_2 \wedge F_2$
to be zero by choosing coordinates as above, we are also again guaranteed  this by 
reducing the tangent bundle to a sub-bundle of rank 3. For instance, we can consider 
the case when our manifold is a product $M^6=X^3 \times Y^3$, with a reduction of the tangent 
bundle as $TM^6=TX^3 \oplus TY^3$, and taking components correspondingly.
%https://math.stackexchange.com/questions/151086/wedge-product-of-differential-form

\paragraph{Case 3: Inclusion of the mixed anomaly term.} 
This case is much more transparently analogous to the heterotic setting indicated above
(provided some inessential complications are set aside). Considering all anomalies at once 
requires the anomaly term to take the form \cite{MMP}\cite{MM}
$$
\int_{M^6} B_2 \wedge Y_4\;,
$$
where $Y_4 \in \Omega^4(M^6; \frak{g})$ is a 4-form with values in the Lie algebra 
$\frak{g}=\bigoplus_I \frak{u}(1)^I$
of abelian structure group $G=U(1)^r$. Here $r$ is the rank, which is less or equal to the 
dimension of the maximal torus of the original non-abelian structure group we have abelianized. 
The physical interpretation of $\frak{g}$ is through the (charge) lattice $\Lambda_\R$, and 
$Y_4$ takes the general form \cite{MMP}\cite{MM}
\(
Y_4=\tfrac{1}{4}ap_1 + \tfrac{1}{2}\sum_I b_{II} c_1^I c_1^I\;,
\label{Y4}
\)
where $p_1$ is the first Pontrjagin class of $M^6$, $c_1^I$ are the Chern-Weil representatives of the first 
Chern class, and the vectors $a, b_{II} \in \Lambda_\R$ are the abelian anomaly coefficients. These vectors 
satisfy explicit intricate constraints, which include, for instance, $6a \cdot b_{II} \in \Z$ and 
$b_{II}^2 + 2a \cdot b_{II} \in 4 \Z$.  In this case, we will have 
\(
\label{gen-dh} 
dH_3=Y_4\;.
\)
Hence, we will have the desired condition, i.e., the closedness of $H_3$, whenever we can arrange 
for $Y_4$ to vanish. Generally, as explained in \cite{PT}\cite{MMP}, explicitly finding configurations 
satisfying the constraints associated with the mixed anomaly cancellation is a nontrivial task. Consequently, we 
will not attempt to specify the corresponding data, especially that our results will not depend on the 
precise numerical factors involved, and instead we will be content that this is possible to arrange,
 ensuring that our constructions below will go through. 

\begin{remark}[Higher structures from anomaly cancellations]
While we are mainly interested in the stronger condition of vanishing of $Y_4$ as a differential form, 
we comment on the structures obtained from requiring the vanishing of the cohomology 
class corresponding to  $Y_4$, i.e., having the expression \eqref{gen-dh} satisfied. 

\item {\bf (i)} Note that conditions of the form $\tfrac{1}{2}p_1 + \alpha_4=0$ are examples of twisted String 
structures \cite{Wa}\cite{SSS2} \cite{SSS3} when $\alpha_4$ is an integral class. Hence we require a twisted
String structure on $M^6$, which is a weakening of the notion of a String structure.
 These aspects will be described in detail, and from a slightly different angle, in \cite{SW3}. 

\item {\bf (ii)} When $\alpha_4$ is not an integral class, we would have a rational twisted String structure. Such 
structures are discussed extensively in \cite{SW1}\cite{SW2}. 

\item {\bf (iii)} Since $\alpha_4$ is proportional to  the product $c_1^2$ of first Chern classes, the twist
takes on a special form. For specific values of the parameters, this is an instance of a String${}^c$ structure 
\cite{CHZ}\cite{tw1}, as a special case of more general constructions \cite{SSS3}\cite{tw2}. 
\end{remark}

%%%%%%%%%%%%%%
\section{Description in twisted (differential) cohomology and twisted (differential) K-theory} 
\label{Sec-K}
%%%%%%%%%%%%%%

In this section we build on the suggestive descriptions in \cref{Sec-cases} and uncover the underlying 
cohomology one step at a time, starting with twisted de Rham cohomology and working our way 
up leading to twisted differential K-theory. 

\medskip
Note that a somewhat similar procedure for the heterotic case led to higher twists \cite{higher}
which were lifted to twisted Morava K-theory and E-theory in \cite{SWe}, 
thereby twisting the description of M-theory and type II string theory via 
the untwisted forms of these generalized cohomology theories \cite{KS1}. 

\medskip
We highlight one interesting effect, nasmely that the twist can arise from the dual field. This  
is, however, similar to what happens in the heterotic case, except that there the dual field is of degree 
seven \cite{higher}. If we are able to apply the self-duality constraint at the level of 
equations of motion but not at the level of action, then we would trade off $*H_3$ for 
$H_3$ and we would get the familiar degree three twisting of the de Rham complex. 
We will, however, keep the discussion general.

\paragraph{The differential complex.} 
Let us first check whether the above discussion leads to  a differential complex. 
We straightforwardly compute the  square of the shifted differential from \eqref{eq-dstarH},
\begin{eqnarray} 
(d-*H_3)^2 &=& d^2 + *H_3 \wedge d  - *H_3 \wedge d - d*H_3 \wedge 
+ *H_3 \wedge *H_3 \wedge
\\ \nonumber
&=& d*H_3 \wedge \;,
\label{dd}
\end{eqnarray} 
whereas for the case of twisting with $H_3$, we would get 
\(
\label{dH}
(d-H_3)^2=dH_3 \wedge.
\)
 From the 6-dimensional Green-Schwarz anomaly cancellation
  \cite{GSW}\cite{GN}\cite{Sag}\cite{Ho}\cite{KMW}\cite{OSTY} \cite{In}\cite{MMP}\cite{MM}, 
  this will be proportional to the topological term 
in \eqref{EOM-Bianchi} or \eqref{gen-dh}. 
However, as explained in each case in \cref{Sec-cases}, we can arrange
for the right-hand-sides to vanish as differential forms. 
We then have $dH_3=0$, so that the 
form $H_3$ lifts to a cohomology class $[H_3]$, which can be a candidate for 
twisting a generalized cohomology theory.

\paragraph{The twisted de Rham complex.}  As we have just seen, in all three cases, the right hand sides of 
the expressions for $dH_3$ (and their dual) can be arranged to be zero. As a consequence, we indeed 
have a twisted differential as well as a corresponding twisted de Rham complex
$$
\xymatrix{ \Omega^2(M^6) \oplus \Omega^4(M^6) \ar[r]^-{d_H} & \Omega^3(M^6) \oplus 
\Omega^5(M^6)},
$$
where $d_H$ is either of the above two differentials, i.e., either $d+H_3\wedge$ or $d+ *H_3 \wedge$. 
We denote by $\Omega_{d_{H}}^i(M^{6}; \R)$ the space of $d_{H}$-closed
differential forms of total degree $i$ on $M^{6}$. 

\paragraph{Duality-symmetric twists.} 
From the action \eqref{originalaction}, we can also consider the variation with respect to $A$
and imposing the Chapline-Manton condition $H_3=F \wedge A$. In this case, we get 
(up to numerical factors) 
$$
dF_4 + h_3 \wedge F_2 + *H_3 \wedge F_2=0\;.
$$
This, together with the Bianchi identity $dF_2=0$, point to a duality-symmetric 
differential of the form 
$$
(d + H_3 \wedge + *H_3 \wedge) \mathcal{F}=0\;.
$$
Indeed, calculating the square of 
$$
d_{H + *H}:=d + H_3 \wedge + *H_3 \wedge
$$
yields zero, $d_{H + *H}^2=0$, ensuring that we can build a 
duality-symmetric de Rham complex. Note that this is analogous to 
a more general (graded nonlinear) case for the M-theory 
fields $G_4$ and $G_7:=*_{11}G_4$ developed in \cite{S1}\cite{tcu}. 

\paragraph{The twisted periodic de Rham complex.} 
Let $R = \R[\![u, u^{-1}]\!]$ be a graded ring where $u$ is a generator 
of degree two. To include periodicity, we use the combination 
\(
\mathcal{F}=u^{-1} F_2 + u^{-2}F_4,
\)
where $u$ is the Bott periodicity generator of degree two. The above expression 
has a uniform degree zero. Let $d_{H_3}=d-u^{-1}H_3$ be the twisted de Rham 
differential of degree one. We denote by $\Omega_{d_{H_3}}^i(M^{6}; \R)[\![ u, u^{-1} ]\!]$ 
the space of $d_{H_3}$-closed differential forms of total degree $i$ on $M^{6}$ in the
periodic case. Here, as above, $H$ refers to either $H_3$ or $*H_3$ or $H_3 + *H_3$. 
 
\paragraph{Twisted periodic de Rham cohomology.} 
The above expression defines an element 
in twisted 2-periodic de Rham cohomology 
$$
[\mathcal{F}] \in H^*(M^6; H_3)[\![ u, u^{-1} ]\!]=
\frac{\ker d_{H_3}}{{\rm Im}\; d_{H_3}}\;,
$$
and similarly for $d_{*H}=d+ *H_3 \wedge$ and $d_{H+*H}=d+ H_3 \wedge + *H_3 \wedge$. 
Now that the similarities are becoming clear, in the sequel we will explicitly illustrate the 
first case (i.e., $H_3$), with the understanding that the 
second and third cases (i.e., $*H_3$ and $H_3 + *H_3$) work analogously. 

\paragraph{Lift to integral cohomology.} We now arrive at a very subtle point. 
A priori, differential cohomology requires curvature with integral periods. \footnote{although there are variants, but they are not what one means generically by the term.} 
In parts of the literature (see \cite{MM} and references therein), it is assumed that  the self-dual fields
in the gravitational and tensor multiplets are described by ordinary (differential) cohomology, 
leaving open the possibility of extension to generalized (differential) cohomology. 
This can be justified by integrality conditions on the fields 
$$
\int_{\Sigma_2} F_2 \;\overset{?}{\in} \; \Z \qquad \text{and} 
\qquad \int_{\Sigma_4} *F_2 \;\overset{?}{\in}\; \Z\;,
$$
for $\Sigma_2$ and $\Sigma_4$ representing appropriate 2-cycles and 4-cycles, respectively.
This would be indicating electric and magnetic flux quantization, so that one can proceed from 
a description via de Rham forms/cohomology to one with integral periods. However, while integrality 
of $F_2$ might be justified, that of $F_4$ might not be as clear. The spectral sequence developed 
in \cite{GS7}\cite{GS8} give precise characterizations in such situations, and more generally for RR fields, but as we have shown above the current system is formally analogous, so that we are able to tap into those detailed results and constructions, which turned out to be quite surprising and subtle for higher degrees. For degree two, it is what one would hope for, i.e., 
$$
[F_2] \!\!\! \mod \Z=0 \;\;\Rightarrow \;\; [F_2] \; \text{is integral} \;\;\Rightarrow \;\; \text{can be refined to a differential class}\; \widehat{F}_2\;. 
$$
The situation for higher degree components, here for $F_4$ (and in the RR case \cite{GS8} for $F_4$ through $F_{10}$). We are not guaranteed that this component a priori has 
integral periods, so that we cannot say that it lifts to a Deligne/differential cohomology class. 
However, the Atiyah-Hirzebruch spectral sequence (AHSS) for twisted differential K-theory 
does give the desired conditions \cite{GS8}, highlighting differences between the flat part and the form part, which we will describe below.

\paragraph{ Lift to twisted K-theory.}
We now investigate whether we can lift the above class $[\mathcal{F}]$ to twisted K-theory. 
In that case,  $u$ would acquire the  interpretation as the Bott generator $u \in K(S^2)$. 
There is a twisted Chern character map 
$$
\xymatrix{
 K(M^6; H_3) \ar[rr]^-{{\rm ch}_{H_3}} && H^{*}(M^6; H_3)[\![u, u^{-1}]\!]
 \ar@/^1.5pc/@{..>}[ll]^{\rm lift?}
}
$$
whose `inverse' is the required lift. 
The lift is a priori obstructed and is governed by the differentials in the Atiyah-Hirzebruch 
spectral sequence (AHSS) for twisted K-theory. We need to check that, on the cohomology classes 
$[F_2]$, $[*F_2]$, and $[H_3]$, corresponding to the form fields $F_2$, $*F_2$, and $H_3$, 
respectively,  we have
\begin{eqnarray} 
{\rm Sq}^3 [F_2] + [H_3] \cup [F_2]&\overset{?}{=}&0 \in H^5(M; \Z)\;,
\label{eq-SqF2}
\\
{\rm Sq}^3 [F_4] + [H_3] \cup [F_4]&\overset{?}{=}&0 \in H^7(M; \Z)\;.
\label{eq-SqF4}
\end{eqnarray}
The second equation certainly holds as it has total degree seven, which is greater than
the dimension of our manifold $M^6$. 
For the first equation, we have ${\rm Sq}^3 [F_2]=0$ by dimension and property of the Steenrod square
of a given degree acting on classes of a lower degree, 
which leaves us with establishing $[H_3] \cup [F_2]=0$. This says that the cup product of the two cohomology 
classes is trivial, so  that at the level of differential forms, $H_3 \wedge F_2$ is an exact form. 
This is indeed compatible with the trivialization arising from the equation $dF_4- H_3 \wedge F_2=0$, 
as a component of \eqref{eq-dstarH}.  

\medskip
Therefore, both of the above equations \eqref{eq-SqF2} and \eqref{eq-SqF4}
hold, and we do indeed have a lift of $[\mathcal{F}]$ to twisted K-theory of $M^6$.
Hence, we have a natural twisted K-theory class on $M^6$ arising from the interaction 
between the Yang-Mills fields from the vector supermultiplet and the $B$-field arising from the 
tensor supermultiplet. Note that torsion classes do arise in global anomaly cancellation 
(see \cite{MM}), but we have obtained a lift without having to explicitly deal with them here.

\paragraph{Lift to twisted differential K-theory.} 
Here we promote the classes $[F_2]$ and $[F_4]$ to differential cohomology classes 
$\widehat{F}_2$ and $\widehat{F}_4$, respectively, twisted by a gerbe with connection 
$\widehat{H}_3$, whose curvature is given by the class $[H_3]$. The lift of elements in 
twisted differential cohomology (or twisted Deligne cohomology) \cite{GS4}\cite{GS5} to 
twisted differential K-theory has been characterized in great detail in \cite{GS7}\cite{GS8}. 
The condition for lifting a degree $n$ differential cohomology class $\mathcal{F}$ 
is generically and schematically (but with subtleties indicated below) given by 
$$
(\widehat{\rm Sq}^3 + \widehat{H}_3 \cup_{\rm DB}) \widehat{\cal F}=0 
\in \mathbb{B}^{n+3}U(1)_\nabla\;,
$$
where 
\begin{itemize} 
\vspace{-2mm} 
\item $\cup_{\rm DB}$ is the Deligne-Beilinson cup product in differential cohomology 
(see \cite{FSS}\cite{FSS2}), 
\vspace{-2mm} 
\item $\widehat{\rm Sq}^3$ is the differential refinement of the Steenrod square 
${\rm Sq}^3$  constructed in \cite{GS2}, 
\vspace{-2mm} 
\item  $\widehat{H}_3$ is the gerbe with connection whose cohomology class is $[H_3]$ 
with differential form representative $H_3$ (see \cite{GS7}),
\vspace{-2mm} 
\item  and  $\mathbb{B}^{n+3}U(1)_\nabla$ is the moduli stack of $U(1)$ $(n+3)$-bundles (or 
abelian $(n+2)$-gerbes) with $(n+1)$-connection $\nabla$ (see \cite{FSSt}\cite{FSS}\cite{FSS2}, 
and \cite{Dur} for a review).
\end{itemize} 
\vspace{-2mm} 
In components, we have for our specific fields,
\begin{eqnarray} 
\widehat{\rm Sq}^3 \widehat{F}_2 + 
\widehat{H}_3 \cup_{\rm DB} \widehat{F}_2&=&0 \in \mathbb{B}^5U(1)_\nabla\;,
\label{eq-SqF2hat}
\\
\widehat{\rm Sq}^3 \widehat{F}_4 + 
\widehat{H}_3 \cup_{\rm DB} \widehat{F}_4&=&0 \in \mathbb{B}^7U(1)_\nabla\;,
\label{eq-SqF4hat}
\end{eqnarray} 
which is a condition of triviality of a composite geometric 4-bundle (or 3-gerbe with 4-connection)
and a composite geometric 6-bundle (or 5-gerbe with 6-connection), respectively. 

\begin{remark}[Subtleties] 
{\bf (i)} There are subtleties related to having the fields one starts with to be integral cohomology classes. As indicated in the discussion of lifting to integral cohomology above, the twisted differential AHSS gives the condition 
$$
a([F_4])= \widehat{\rm Sq}^3 (\widehat{F}_2)\;,
$$
where $a: \Omega^4(M^6) \longrightarrow \widehat{H}^5(M^6; \Z)$ is the inclusion of the 5-bundles with globally defined connection 4-forms. Since we do not have $F_6$, the next expression we have is 
$$
a(F_6)=0=j_2 \overline{\rm Sq}^2 (F_4)
$$
where $\overline{\rm Sq}^2$ is an operation which is well-defined on half-integral classes
and only well-defined on the $E_3$-page of the twisted differential AHSS.  This cannot be 
modified to $\widehat{\rm Sq}^3$ unless $[F_4]$ is in fact integral, in which case it does describe 
a Deligne class, and the above equation reduces to $\widehat{\rm Sq}^3(\widehat{F}_4)$. The twisted 
equation, instead of  \eqref{eq-SqF2hat}, then looks like 
\(
\widehat{\rm Sq}^3 (\widehat{F}_2) - \widehat{H}_3 \cup \widehat{F}_2 = a(F_4)\;,
\label{alternative-eq-SqF2hat}
\)
which, at the level of curvatures, gives $dF_4 + H_3 \wedge F_2=0$.

\item {\bf (ii)} 
One point to highlight here is that, without the addition of the anomaly term in the action, the 
exponentiated action is manifestly anomalous just by virtue of the K-theoretic interpretation. 
Indeed, we just saw by the above discussion that $F_4$ does not necessarily have integral periods. Varying the K-theory class of the $F$'s in the action will thus result in not necessarily 
integral jumps on the moduli space of fields, leading to an anomaly. The term $B_2\wedge Y_4$ is meant to fix this, so that the combination is integral, i.e., what the twisted Chern character 
is meant to do.

\end{remark}

Hence, by the above discussion that builds on the 
general results in \cite{GS2}\cite{GS8}, we have that the validity of 
\eqref{eq-SqF2} and \eqref{eq-SqF4} ensure, respectively, the validity of expressions
\eqref{eq-SqF2hat} and \eqref{eq-SqF4hat} (or \eqref{alternative-eq-SqF2hat} in the non-integral case).

\medskip
Subject to the above discussion, we collect together our results thus far in the following: 

\begin{theorem}[Characterization via twisted cohomology] 
Consider the Yang-Mills fields in six-dimensional ${\cal N}=(1,0)$ theory
with a tensor multiplet, including the B-field $H_3$. When abelianized (and subject to constraints 
from \cref{Sec-cases}), we have for the total field $\mathcal{F}:=u^{-1} F_2 + u^{-2} *F_2=u^{-1} F_2 + u^{-2} F_4$:
\begin{enumerate}[{\bf (i)}]
\vspace{-1mm} 
\item  (Twisted de Rham) The class of $\mathcal{F}$ is $[\mathcal{F}] \in H^*(M^6; H)[\![ u, u^{-1}]\!]$, 
where the twist is given either by the closed form $H_3$, its Hodge dual $*H_3$, or their combination 
$H_3 + *H_3$. 
\vspace{-2mm} 
\item (Twisted K-theory) The class $[\mathcal{F}]$ lifts to a class $\mathcal{F}_{K_H}$ in 
twisted K-theory $K(M^6; [H])$, where the twist $[H]$ is given by the class $[H_3]$, its dual 
$[*H_3]$, or the combination $[H_3 + *H_3]=[H_3]+ [*H_3]$.
\vspace{-2mm} 
\item (Twisted differential K-theory) The class $\mathcal{F}_{K_H}$ lifts to a class $\widehat{\mathcal{F}}$ in 
twisted differential K-theory $\widehat{K}(M^6; \widehat{H})$, where $\widehat{H}$ is the differential 
cohomology class whose curvature class is either $[H_3]$, $[*H_3]$, or $[H_3+*H_3]$, 
where in the non-integral case we require condition \eqref{alternative-eq-SqF2hat} to hold.
\end{enumerate} 
\end{theorem} 

%%%%%%%%%%%%%%%%%%
\section{Description using untwisted higher cohomology theories}
\label{Sec-un}  
%%%%%%%%%%%%%%%%%%

The discussion in this section applies to both $\mathcal{N}=(1,0)$ and $\mathcal{N}=(2,0)$ 
theories and, to a large extent, both type IIA and type IIB string  theories. The analysis is 
also preliminary as a full description requires dealing with torsion. On the one had this would be very 
rich, while on the other hand it would be highly nontrivial in the absence of a good understanding 
of those aspects of the six-dimensional theories, and hence such a study would  inevitably need 
to go beyond the scope fo this note. 

\medskip
We discuss whether another interpretation of (the refinement of) $H_3$, not as a twist, but as an 
invertible element, i.e., a higher (categorical/cohomological) degree analog of a line bundle. 
Mathematically, this would be an element in the Picard group of the corresponding cohomology 
theory -- see \cite{SWe}\cite{LSW}\cite{GS7} for extensive discussions. 

\medskip
Issues with incompatibility of twisted K-theory and S-duality in ten dimension \cite{KS2} have 
led to the proposal that the Ramond-Ramond (RR) fields might live in an untwisted cohomology 
theory, where the Neveu-Schwarz (NS) field is considered simply as an invertible element in that 
theory. Note that while S-duality in six dimensions does not necessarily lead to such incompatibilities 
(yet there is self-duality issues), the putative interpretation nevertheless is a possibility that could 
be entertained, especially in the absence of a proper understanding of the $\mathcal{N}=(1,0)$ 
and $\mathcal{N}=(2,0)$ theories. 
Note that S-duality in six dimensions can be treated in a novel way via noncommutative 
geometry \cite{MS}. 

\medskip
In the case of K-theory, line bundles are invertible elements of the theory. They are classified 
by the first Chern class $c_1$ in cohomological degree two and correspond to maps from 
the classifying space $BU(1)\simeq \mathbb{C} P^\infty$, which has the same homotopy 
type as the Eilenberg-MacLane space $K(\Z, 2)$,  to the K-theory spectrum $\mathcal{K}$. 
Such a map is reminiscent of the map $K(\Z, 3) \to B{\rm GL}_1(\mathcal{K})$ defining 
the determinental twist of the theory. 

\medskip
Next we look at the case at hand, which is for $H_3$. Topologically, we are in cohomological 
degree three, so we seek the existence of maps from $K(\Z, 3)$  to a spectrum $\mathcal{E}$
corresponding to some generalized cohomology theory $E^\bullet(M^6)$. Such a maps can 
be related to maps arising from the twists of the spectrum $\mathcal{E}$, 
$$
\text{universal\;twist}: \;\;K(\Z, 4) \longrightarrow B{\rm GL}_1(\mathcal{E})\;.
$$
So now, our task is  reduced to identifying natural candidate spectra $\mathcal{E}$ which admit
twists of degree four. 
\footnote{Note that, in practice, for the purpose of twisting theories, the logic is the other way around. 
However, since for our purposes we are trying to detect existing theories rather than provide new twists, 
and our focus is the current physics application, we tap into twists that already exist to deduce the (a priori)
underlying maps. Alternatively, such maps can be deduced in a case by case basis, as we also to some extent 
do below, but their existence depends on specifics about the theory (String orientation, viewing gerbes as instances
of 2-vector bundles, etc.). Note also that we are in the stable setting, so that looping a map 
is guaranteed to give a nontrivial map. 
} 
We indeed identify four cohomology theories corresponding to such spectra,
and provide arguments for the  relevance of each. These are 
\begin{enumerate}[{\bf (i)}]
\vspace{-2mm} 
\item topological modular forms (tmf);
\vspace{-2mm}
\item Morava K-theory $K(2)$;
\vspace{-2mm}
\item Morava E-theory $E(2)$; 
\vspace{-2mm}
\item Algebraic K-theory of the topological complex K-theory spectrum $K_{\rm alg}(\mathcal{K})$.
\end{enumerate} 
Note that the first three \cite{AHS}\cite{JW} arise in chromatic homotopy theory as being at 
chromatic level two, while the fourth is at categorical level two (see \cite{LSW}) and is proposed 
to be a form of elliptic cohomology \cite{BDR}.

\medskip
Note that this is somewhat analogous to viewing a closed form in two different ways, leading to 
two different interpretations, namely as either a flat $n$-connection in 
$n$ dimensions or as an $n$-curvature in $n+1$ dimensions (see \cite{sigma}\cite{FSS}),
which involves a transgression or a shift in degree by one. In the current case, we have
a shift in degree as well, instead of viewing a closed form as corresponding to a map 
from $K(\Z, n) \to B{\rm GL}_1(\mathcal{E}))$ corresponding to the twist, we view 
it as a map $K(\Z, n) \to \mathcal{E}$, which is off by one degree due to delooping.

\medskip
This is for the M5-brane theory. Here we do something analogous, but with different 
approach and tools.

\paragraph{From twisted K-theory to topological modular forms.} 
 The determinantal twists of K-theory can be included as ``elliptic line bundles" 
in the theory of topological modular forms \cite{Doug}\cite{KS2}. 
\footnote{There are various versions of this theory, including periodic TMF and connective tmf. 
Here we will not make a big distinction, as the twists and corresponding maps hold equally for both.
The same holds for the K-theory spectrum appearing in the algebraic K-theory of that spectrum 
$K_{\rm alg}(\mathcal{K})$.}
There is a map 
$K(\Z,3)\to {\rm tmf}$ coming from the String  orientation. A twist of K-theory of the form 
$M^6 \to K(\Z,3)$ gives rise to an  element of ${\rm tmf}(X)$ by composition $M^6 \to K(\Z,3) \to {\rm tmf}$.  
The map $M^6 \to K(\Z, 3)$ corresponds to a $BS^1$ bundle over $M^6$ which can
be described geometrically as a 1-dimensional 2-vector bundle, hence a
higher notion of a line bundle. This is generalized considerably in \cite{LSW}. 
Note that this map can also be seen from twisted tmf and the existence of a map 
$K(\Z, 4) \to B{\rm GL}_1({\rm tmf})$ \cite{ABG}, building on proposals of the author 
(see \cite{S1}\cite{tcu}). 

\medskip
From modularity and F-theory \cite{KS3} and S-duality in type IIB \cite{KS2}, it was proposed that tmf
should play an important role in formulating type II theories. 
In \cite{tcu} it was argued that topological modular forms (tmf) \cite{AHS} plays a role for 
M-branes and indeed twisted string structures have been identified on the worldvolume 
\cite{tcu}\cite{tw1}\cite{tw2}. Recently, on the field theory side, which is more directly
relevant to our discussion here, there has been an interesting 
connection to tmf \cite{GPPV}.

\medskip
\paragraph{Relation to Morava K-theory and E-theory.} 
The investigations in \cite{higher} have led to proposing that Morava K-theory $K(2)$ 
and E-theory $E(2)$, which had proven their value in anomaly cancellation in M-theory 
and type IIA string theory \cite{KS1}, can be twisted by higher degree twists than those
appearing traditionally in string theory using K-theory. Indeed, this was proved in 
vast generality in \cite{SWe} at every chromatic level and for any prime $p$. The
twists are given by maps 
$$
K(\Z, 4) \longrightarrow B{\rm GL}_1(K(2)) 
\qquad \text{and} \qquad
K(\Z, 4) \longrightarrow B{\rm GL}_1(E(2)) 
$$
Correspondingly, we have maps $K(\Z, 3) \to K(2)$ and $K(\Z, 3) \to E(2)$, interpreting elements
of cohomological degree three as invertible elements in these theories. Thus, the integral cohomology
class $[H]$ corresponding to either $H_3$, $*H_3$, or $H_3 + *H_3$ can be viewed as an invertible 
element in $K(2)$ or $E(2)$, i.e., these are the line bundles in these theories. 

%\medskip
%Integral lifts of Morava

\medskip
\paragraph{Relation to algebraic K-theory of the topological K-theory spectrum.} 
This is perhaps the theory that might most easily and explicitly connected to physics,
as it is the Grothendieck group of (not necessarily abelian) 2-vector bundles, hence 
connecting to the (abelian) 2-bundle description via higher moduli stacks (see \cite{E8-stack} 
for the case of M-theory reducing to the heterotic B-field, and \cite{Dur} for a gentle review). 
It was shown in \cite{LSW} that the theory $K_{\rm alg}(\mathcal{K})$, being at categorical 
level two, admits twists of cohomological degree four, hence there are maps 
$$
K(\Z, 4) \longrightarrow B{\rm GL}_1(K_{\rm alg}(\mathcal{K}))\;.
$$
Indeed, the corresponding map $K(\Z, 3) \to K_{\rm alg}(\mathcal{K})$ is the 
construction that views a gerbe as a 2-vector bundle (see \cite{BDR}). Hence, the classes 
$[H]$ can be viewed as invertible elements, i.e., 2-line bundles, in the form of elliptic 
cohomology $K_{\rm alg}(\mathcal{K})$.

\begin{prop}[Description via untwisted `higher' generalized cohomology] 
\label{Prop}
The tensor multiplet field $H$, in the $\mathcal{N}=(2,0)$ theory and in the 
$\mathcal{N}=(1,0)$ theory with vanishing anomaly,  lifts to an invertible 
element in untwisted  
\begin{enumerate}[{\bf (i)}]
 \vspace{-2mm}
\item Topological Modular Forms ${\rm tmf}(M^6)$ via $M^6 \to K(\Z, 3) \to {\rm tmf}$;
 \vspace{-2mm}
\item Morava K-theory $K(2)(M^6)$ via $M^6 \to K(\Z, 3) \to K(2)$;
 \vspace{-2mm}
\item Morava E-theory $E(2)(M^6)$, via $M^6 \to K(\Z, 3) \to E(2)$;
 \vspace{-2mm}
\item Algebraic K-theory of topological complex K-theory $K_{\rm alg}(KU)(M^6)$, via $M^6 \to K(\Z, 3) \to K_{\rm alg}(KU)$.
%by viewing the gerbe associated to $H_3$ as a virtual 2-vector bundle. 
\end{enumerate} 
\end{prop}

\medskip
\begin{remark}[Interpretations and extensions] We highlight the following on the statements in the Prop. 
\ref{Prop}:
\vspace{-1mm} 
\item {\bf (i)} 
In principle, the above statements extends to the differential case. While it is abstractly mathematically 
guaranteed that differential versions of the above theories exist, construction of such 
has only been carried out for two of them, namely the Morava theories \cite{GS6}. 
For these, the first (and only a priori relevant) 
differential in the AHSS for these theories has been identified as $\widehat{Q}_2$, the 
differential refinement of the second Milnor primitive $Q_2$ appearing in the AHSS 
in the topological case \cite{KS1}, and built out of the differential Steerod square $\widehat{\rm Sq}^3$
constructed in \cite{GS2}. Viewing this as an obstruction, we see that it vanishes by dimension, so that
we certainly have a lift to Morava K-theory/E-theory as well as their differential refinements.
\vspace{-1mm} 
\item {\bf (ii)} The discussion on tmf can be viewed as providing further topological support to the 
interesting proposals mostly on the modular forms part in \cite{GPPV}. We hope that combining various
angles would lead to a more complete and nicely emerging picture. 
\vspace{-1mm} 
\item {\bf (iii)} Note that these theories require working with one prime at a time. 
However, there are integral versions, for instance for Morava K-theory \cite{KS1}\cite{SWe}.
\vspace{-1mm} 
\item {\bf (iv)} For a full consideration and to unpack the consequences explicitly, 
we would need to include torsion, including and going beyond
the considerations \cite{MMP}.  This deserves to be dealt with elsewhere. 
\end{remark}

 \vspace{.2cm}
\noindent {\large \bf Acknowledgement.} 
The author would like to thank Dan Grady for very useful discussions and Eric Sharpe and 
Seok Kim for useful comments. This work was done at the Aspen Center for Physics, 
which is supported by National Science Foundation grant PHY-1607611. This work 
was also partially supported by a grant from the Simons Foundation.

%%%%%%%%%%%%%%

 \medskip
\noindent Hisham Sati, {\it Mathematics, Division of Science, New York University Abu Dhabi, UAE.}


\begin{thebibliography}{1}
%%%%%%%%%%%%

\bibitem[ABG10]{ABG} 
M. Ando, A. J. Blumberg, and D. J. Gepner, 
{\it Twists of K-theory and TMF}, 
Superstrings, geometry, topology, and $C^\ast$-algebras, 27--63, 
Proc. Sympos. Pure Math., 81, Amer. Math. Soc., Providence, RI, 2010,
[\href{https://arxiv.org/abs/1002.3004}{\tt arXiv:1002.3004}] [{\tt math.AT}].

\vspace{-2mm} 
\bibitem[AHS01]{AHS}
M. Ando, M. J. Hopkins, and N. P. Strickland, 
\href{http://web.math.rochester.edu/people/faculty/doug/otherpapers/musix.pdf}
{\it Elliptic spectra, the Witten
genus and the theorem of the cube}, Invent. Math.  {\bf 146} (2001), 595-687.


\vspace{-2mm} 
\bibitem[BDR04]{BDR}
N. A. Baas,  B. I. Dundas, and J. Rognes, 
{\it Two-vector bundles and forms of elliptic cohomology},
Topology, geometry and quantum field theory,
Cambridge Univ. Press, Cambridge, 2004, 
[\href{https://arxiv.org/abs/1002.3004}{\tt	arXiv:math/0306027}] [{\tt math.AT}]. 
 %               URL = {http://dx.doi.org/10.1017/CBO9780511526398.005},


\vspace{-2mm} 
\bibitem[Bh15]{Bh} 
 L. Bhardwaj, {\it Classification of $6d$ $N = (1, 0)$ gauge theories}, J. High Energy Phys. {\bf 11} (2015) 002, 
 \newline
 [\href{https://arxiv.org/abs/1502.06594}{\tt arXiv:1502.06594}][{\tt hep-th}].

\vspace{-2mm} 
 \bibitem[BdRdWvN82]{BddN}
E. Bergshoeff, M. de Roo, B. de Wit, and P. van Nieuwenhuizen, 
{\it Ten-dimensional Maxwell-Einstein supergravity, its currents, and the issue of its 
auxiliary fields}, Nucl. Phys. {\bf B 195} (1982), 97-136.

 
 \vspace{-2mm} 
 \bibitem[CM83]{CM}
G. F. Chapline and N. S. Manton,
{\it  Unification of Yang-Mills theory and supergravity in ten-dimensions}, 
Phys. Lett. {\bf B 120} (1983) 105-109.

\vspace{-2mm} 
\bibitem[CHZ11]{CHZ}
Q. Chen, F. Han, and W. Zhang,
{\it Generalized Witten Genus and Vanishing Theorems}, 
J. Diff. Geom. {\bf 88} (2011), 1-39,
[\href{https://arxiv.org/abs/arXiv:1003.2325}{\tt arXiv:1003.2325}] [{\tt math.DG}].

\vspace{-2mm} 
\bibitem[Do06]{Doug}
C. Douglas, 
{\it On the twisted $K$-homology of simple Lie groups},
Topology {\bf 45} (2006), 955-988,  \newline
[\href{https://arxiv.org/abs/math.AT/0402082}{\tt arXiv:math.AT/0402082}].

\vspace{-2mm} 
\bibitem[FRS98]{FRS}
S. Ferrara, F. Riccioni and A. Sagnotti, Tensor and vector multiplets in 
six-dimensional supergravity,
Nucl. Phys. {\bf B 519} (1998) 115-140, 
[\href{https://arxiv.org/abs/hep-th/9711059}{\tt arXiv:hep-th/9711059}].

\vspace{-2mm} 
\bibitem[FSS13]{FSS}
D.~Fiorenza, H. Sati, and U.~Schreiber, 
{\it Extended higher cup-product Chern-Simons theory}, 
J. Geom. Phys. {\bf 74} (2013), 130--163,
[\href{https://arxiv.org/abs/1207.5449}{\tt arXiv:1207.5449}] [{\tt hep-th}]. 
	

\vspace{-2mm} 
\bibitem[FSS15a]{FSS2}
D.~Fiorenza, H. Sati, and U.~Schreiber, 
{\it A Higher stacky perspective on Chern-Simons theory}, 
Mathematical Aspects of Quantum Field Theories (D. 
      Calaque and T. Strobl eds.), Springer, Berlin (2015),
[\href{https://arxiv.org/abs/1301.2580}{\tt arXiv:1301.2580}] [{\tt hep-th}].

\vspace{-2mm} 
\bibitem[FSS15b]{E8-stack}
D. Fiorenza, H. Sati, and U. Schreiber,
{\it The $E_8$ moduli 3-stack of the $C$-field},
Commun. Math. Phys. {\bf 333} (2015),  117-151,
[\href{https://arxiv.org/abs/1202.2455}{\tt arXiv:1202.2455}][{\tt hep-th}].


\vspace{-2mm} 
\bibitem[FSS19a]{Dur}
D. Fiorenza, H. Sati,  and U. Schreiber,
{\it The rational higher structure of M-theory},
Proc. LMS-EPSRC Durham Symposium
{\it Higher Structures in M-Theory}, August 2018,
Fortsch. der Phys., 2019, \newline
[\href{https://doi.org/10.1002/prop.201910017}{\tt doi:10.1002/prop.201910017}]
[\href{https://arxiv.org/abs/1903.02834}{\tt arXiv:1903.02834}][{\tt hep-th}].


\vspace{-2mm} 
\bibitem[FSS19b]{FSS9} 
D. Fiorenza, H. Sati,  and U. Schreiber,
{\it Super-exceptional geometry: origin of heterotic M-theory and super-exceptional embedding construction of M5}, 
[\href{https://arxiv.org/abs/1908.00042}{\tt	arXiv:1908.00042}] [{\tt hep-th}].


\vspace{-2mm} 
\bibitem[FSSt12]{FSSt}
D. Fiorenza, U. Schreiber, and J. Stasheff, 
{\it {\v C}ech cocycles for differential characteristic classes -- An 
infinity-Lie theoretic construction},
Adv. Theor. Math. Phys. {\bf 16} (2012), 149--250,  
[\href{https://arxiv.org/abs/1011.4735}{\tt  arXiv:1011.4735}] [{\tt math.AT}].



\vspace{-2mm} 
\bibitem[Fr00]{Fr}
D. S. Freed, {\it Dirac charge quantization and generalized differential cohomology}, 
with an Appendix with M. Hopkins,  Surv. Differ. Geom. {\bf 7}, pp. 129-194, Int. Press, Somerville, 
MA, 2000, 	[\href{https://arxiv.org/abs/hep-th/0011220}{\tt arXiv:hep-th/0011220}]. 

 \vspace{-2mm} 
 \bibitem[GS17]{GS6}
D. Grady and H. Sati, {\it Spectral sequence in smooth generalized cohomology}, 
Algebr. Geom. Top. {\bf 17} (2017), 2357-2412,
[\href{https://arxiv.org/abs/1605.03444}{\tt arXiv:1605.03444}] [{\tt math.AT}].


\vspace{-2mm} 
 \bibitem[GS18a]{GS2}
D. Grady and H. Sati, {\it Primary operations in differential cohomology}, 	
Adv. Math. {\bf 335} (2018), 519-562, \newline
[\href{https://arxiv.org/abs/1604.05988}{\tt arXiv:1604.05988}] [{\tt math.AT}].
 
 \vspace{-2mm} 
\bibitem[GS18b]{GS4}
D. Grady and H. Sati,
{\it Twisted smooth Deligne cohomology},
Ann. Global Analysis Geom. {\bf 53} (2018), 445-466,  
[\href{https://arxiv.org/abs/1706.02742}{\tt arXiv:1706.02742}] [{\tt math.DG}].


 \vspace{-2mm} 
 \bibitem[GS19a]{GS5}
D. Grady and H. Sati,
{\it Higher-twisted periodic smooth Deligne cohomology}, 
Homology, Homotopy and Appl. {\bf 21} (2019)
129-159,	
[\href{https://arxiv.org/abs/1712.05971}{\tt arXiv:1712.05971}] [{\tt math.DG}].



\vspace{-2mm} 
 \bibitem[GS19b]{GS7}
D. Grady and H. Sati,
{\it Twisted differential generalized cohomology theories and their Atiyah-Hirzebruch spectral sequence},
Alg. Geom. Topol. (2019),
[\href{https://arxiv.org/abs/1711.06650}{\tt arXiv:1711.06650}][{\tt math.AT}].

 
 \vspace{-2mm} 
 \bibitem[GS19c]{GS8}
D. Grady and H. Sati,
{\it Ramond-Ramond fields and twisted differential K-theory},
[\href{https://arxiv.org/abs/1903.08843}{\tt arXiv:1903.08843}] [{\tt hep-th}].

 
 \vspace{-2mm} 
 \bibitem[GSW85]{GSW}
 M. B. Green, J. H. Schwarz and P. C. West, 
{\it Anomaly Free Chiral Theories in Six-Dimensions}, 
 Nucl. Phys. {\bf B 254} (1985), 327-348. 
 
 \vspace{-2mm} 
 \bibitem[GPPV18]{GPPV} 
S. Gukov, D. Pei, P. Putrov, and C. Vafa,
{\it 4-manifolds and topological modular forms}, \newline
[\href{https://arxiv.org/abs/1811.07884}{\tt arXiv:1811.07884}] [{\tt hep-th}].

\vspace{-2mm} 
\bibitem[GSS11]{GSS} 
 M. Gunaydin, H. Samtleben and E. Sezgin, {\it On the Magical Supergravities in Six
Dimensions}, Nucl. Phys. {\bf B 848} (2011) 62-89,
[\href{https://arxiv.org/abs/1012.1818}{\tt arXiv:1012.1818}] [{\tt hep-th}].

\vspace{-2mm} 
\bibitem[HZ98]{HZ} 
 A. Hanany and A. Zaffaroni, {\it Branes and six-dimensional supersymmetric theories},
Nucl. Phys. {\bf B 529} (1998) 180-206, 
[\href{https://arxiv.org/abs/hep-th/9712145}{\tt arXiv:hep-th/9712145}].


\vspace{-2mm} 
\bibitem[HMV14]{HMV}
 J. J. Heckman, D. R. Morrison, and C. Vafa, {\it On the Classification of 6D SCFTs and
Generalized ADE Orbifolds}, J. High Energy Phys. {\bf 05} (2014) 028,
[\href{https://arxiv.org/abs/1312.5746}{\tt arXiv:1312.5746}][{\tt hep-th}].

\vspace{-2mm} 
\bibitem[HMRV15]{HMRV}
 J. J. Heckman, D. R. Morrison, T. Rudelius, and C. Vafa, {\it Atomic Classification of 6D
SCFTs}, Fortsch. Phys. {\bf 63} (2015) 468-530,
[\href{https://arxiv.org/abs/1502.05405}{\tt arXiv:1502.05405}][{\tt hep-th}].

 \vspace{-2mm} 
 \bibitem[Ho06]{Ho} 
G. Honecker, {\it Massive U(1)s and heterotic five-branes on K3}, 
 Nucl. Phys. {\bf B 748} (2006), 126-148,  \newline
 [\href{https://arxiv.org/abs/hep-th/0602101}{\tt arXiv:hep-th/0602101}].
 
\vspace{-2mm} 
\bibitem[In14]{In} 
K. Intriligator, 
{\it 6d, ${\cal N}=(1,0)$ Coulomb Branch Anomaly Matching}, 
J. High Energy Phys. {\bf 10} (2014) 162,
 [\href{https://arxiv.org/abs/1408.6745}{\tt arXiv:1408.6745}][{\tt hep-th}].

\vspace{-2mm} 
\bibitem[JW75]{JW}
D. C. Johnson and W. S. Wilson, {\it BP operations and Morava's extraordinary K-theories},
 Math. Z. {\bf 144} (1975),  55-75.

\vspace{-2mm} 
\bibitem[KKP16]{KKP}
H.-C. Kim, S. Kim, and J. Park,
{\it 6d strings from new chiral gauge theories}, 
 [\href{https://arxiv.org/abs/1608.03919}{\tt arXiv:1608.03919}] [{\tt hep-th}].

\vspace{-2mm} 
\bibitem[KS04]{KS1}
I.~Kriz and H.~Sati, {\it M Theory, type IIA superstrings, and
elliptic cohomology}, Adv. Theor. Math. Phys. {\bf 8} (2004) 345-395,
[\href{https://arxiv.org/abs/hep-th/0404013}{\tt  arXiv:hep-th/0404013}].

\vspace{-2mm} 
\bibitem[KS05a]{KS2}
I.~Kriz and H.~Sati, {\it Type IIB string theory, S-duality and
generalized cohomology}, Nucl. Phys. {\bf B 715} (2005) 639--664, 
[\href{https://arxiv.org/abs/hep-th/0410293}{\tt arXiv:hep-th/0410293}].

\vspace{-2mm} 
\bibitem[KS05b]{KS3}
I.~Kriz and H.~Sati, {\it Type II string theory and modularity}, 
	J. High Energy Phys.  {\bf 0508} (2005) 038, 
[\href{https://arxiv.org/abs/hep-th/0501060}{\tt 	arXiv:hep-th/0501060}].

 \vspace{-2mm} 
 \bibitem[KMW10]{KMW} 
 V. Kumar, D. R. Morrison, and W. Taylor, 
 {\it Global aspects of the space of 6D N = 1 supergravities}, 
 J. High Energy Phys. {\bf 1011} (2010) 118, 
 [\href{https://arxiv.org/abs/1008.1062}{\tt   arXiv:1008.1062}][{\tt hep-th}].

\vspace{-2mm} 
 \bibitem[LRW18]{LRW}
 S.-J. Lee, D. Regalado, and T. Weigand ,
{\it 6d SCFTs and U(1) Flavour Symmetries}, 
J. High Energy Phys. {\bf 1811} (2018) 147,
 [\href{https://arxiv.org/abs/1803.07998}{\tt  arXiv:1803.07998}] [{\tt hep-th}].

\vspace{-2mm} 
\bibitem[LSW19]{LSW}
J. A. Lind, H. Sati, and C. Westerland,
{\it Twisted iterated algebraic K-theory and topological 
T-duality for sphere bundles}, Ann. K-theory (2019), 
 [\href{https://arxiv.org/abs/1601.06285}{\tt arXiv:1601.06285}] [{\tt math.AT}]. 

\vspace{-2mm} 
\bibitem[MS15]{MS}
 V. Mathai and H. Sati,
{\it Higher abelian gauge theory associated to gerbes on noncommutative 
deformed M5-branes and S-duality},
J. Geom. Phys. {\bf 92} (2015), 240-251,
[\href{https://arxiv.org/abs/1404.2257}{\tt arXiv:1404.2257}][{\tt hep-th}].
%doi: 10.1016/j.geomphys.2015.02.019 

 \vspace{-2mm} 
\bibitem[MM18]{MM} 
S. Monnier and G. W. Moore,
{\it Remarks on the Green-Schwarz terms of six-dimensional supergravity theories}, 
[\href{https://arxiv.org/abs/1808.01334}{\tt arXiv:1808.01334}] [{\tt hep-th}].

\vspace{-2mm} 
  \bibitem[MMP18]{MMP} 
 S. Monnier, G. W. Moore, and D. S. Park, 
 {\it Quantization of anomaly coefficients in 6D $N = (1,0)$ supergravity},
J. High Energy Phys. {\bf 1802} (2018) 020, 
[\href{https://arxiv.org/abs/1711.04777}{\tt arXiv:1711.04777}] [{\tt hep-th}].

\vspace{-2mm} 
  \bibitem[Mo12]{Moore} 
G. Moore,
{\it Applications of the six-dimensional $(2,0)$ theories to Physical Mathematics},
Felix Klein lectures, Bonn 2012,
[\url{http://www.physics.rutgers.edu/~gmoore/FelixKleinLectureNotes.pdf}]

 \vspace{-2mm} 
 \bibitem[NG86]{GN} 
 H. Nishino and S. James Gates, Jr., 
 {\it Dual Versions of Higher Dimensional Supergravities and Anomaly 
 Cancellations in Lower Dimensions},  Nucl. Phys. {\bf B 268} (1986), 532-542. 

\vspace{-2mm} 
 \bibitem[NS97]{NS} 
 H. Nishino and E. Sezgin, {\it New couplings of six-dimensional supergravity}, 
 Nucl. Phys. {\bf B 505} (1997) 497-516,
[\href{https://arxiv.org/abs/hep-th/9703075}{\tt  arXiv:hep-th/9703075}].

 
\vspace{-2mm} 
 \bibitem[OSTY14]{OSTY} 
 K. Ohmori, H. Shimizu, Y. Tachikawa and K. Yonekura, {\it Anomaly polynomial of
general 6d SCFTs},  \href{https://doi.org/10.1093/ptep/ptu140}{Prog. Theor. Exp. Phys. {\bf 2014} (2014), 103B07}, 
%\url{https://doi.org/10.1093/ptep/ptu140},
[\href{https://arxiv.org/abs/1408.5572}{\tt arXiv:1408.5572}] [{\tt hep-th}].

\vspace{-2mm} 
\bibitem[PT12]{PT}
D. S. Park and W. Taylor,
{\it Constraints on 6D Supergravity Theories with Abelian Gauge Symmetry}, 
J. High Energy Phys. {\bf 1201} (2012) 141,
 [\href{https://arxiv.org/abs/1110.5916}{\tt arXiv:1110.5916}] [{\tt hep-th}]. 
 
 \vspace{-2mm} 
\bibitem[Ri00]{Ric} 
F. Riccioni, {\it Abelian vector multiplets in six-dimensional supergravity}, 
Phys. Lett. {\bf B 474} (2000) 79-84,
 [\href{https://arxiv.org/abs/hep-th/9910246}{\tt arXiv:hep-th/9910246}].

\vspace{-2mm} 
\bibitem[RS98]{RS}
F. Riccioni and A. Sagnotti, {\it Some properties of tensor multiplets in six-dimensional
supergravity}, Nucl. Phys. Proc. Suppl. {\bf 67} (1998) 68-73,
 [\href{https://arxiv.org/abs/hep-th/9711077}{\tt  arXiv:hep-th/9711077}].

\vspace{-2mm}  
 \bibitem[Sg92]{Sag} 
 A. Sagnotti,
{\it  A Note on the Green-Schwarz Mechanism in Open-String Theories}, 
Phys. Lett. {\bf B 294} (1992), 196-203,
 [\href{https://arxiv.org/abs/hep-th/9210127}{\tt  arXiv:hep-th/9210127}].

\vspace{-2mm} 
\bibitem[SSW11]{SSW1}
  H. Samtleben, E. Sezgin, and R. Wimmer, {\it (1,0) superconformal 
  models in six dimensions}, J. High Energy Phys. {\bf 1112} (2011) 062,
    [\href{https://arxiv.org/abs/1108.4060}{\tt arXiv:1108.4060}] [{\tt hep-th}].
  
  \vspace{-2mm}
  \bibitem[Sa06]{S1}
  H. Sati, 
{\it  Duality symmetry and the form fields of M-theory}, 
J. High Energy Phys. {\bf 0606} (2006) 062, 
 [\href{https://arxiv.org/abs/hep-th/0701232}{\tt arXiv:hep-th/0509046}].
  
 \vspace{-2mm} 
  \bibitem[Sa09]{higher}
 H. Sati,   {\it A Higher Twist in String Theory}, 
	J. Geom. Phys. {\bf 59} (2009), 369-373,
%DOI:	10.1016/j.geomphys.2008.11.009
 [\href{https://arxiv.org/abs/hep-th/0701232}{\tt	arXiv:hep-th/0701232}].
 

 \vspace{-2mm} 
 \bibitem[Sa10]{tcu} 
 H. Sati, {\it Geometric and topological structures related to M-branes},
 Superstrings, geometry, topology, and $C^*$-algebras, 181--236, Proc. Sympos. Pure Math., 81, 
 Amer. Math. Soc., Providence, RI, 2010, 	
  [\href{https://arxiv.org/abs/1001.5020}{\tt arXiv:1001.5020}] [{\tt math.DG}].
 
 \vspace{-2mm} 
 \bibitem[Sa11b]{tw1}
 H. Sati, {\it Geometric and topological structures related to M-branes II: 
 Twisted String and ${\rm String}^c$
 structures}, J. Australian Math. Soc. {\bf 90} (2011), 93-108,
 [\href{https://arxiv.org/abs/1007.5419}{\tt arXiv:1007.5419}] [{\tt hep-th}]. 


  \vspace{-2mm} 
 \bibitem[Sa11a]{E8}
H. Sati, {\it Anomalies of $E_8$ gauge theory on string manifolds}, 
Int. J. Mod. Phys. {\bf A 26} (2011), no. 13, 2177--2197,
 [\href{https://arxiv.org/abs/0807.4940}{\tt arXiv:0807.4940}] [{\tt hep-th}].
 
 
 
 \vspace{-2mm} 
\bibitem[Sa11b]{tw2}
H. Sati, {\it Twisted topological structures related to M-branes}, 
Int. J. Geom. Methods Mod. Phys. {\bf 8} (2011), no. 5, 1097--1116,
[\href{https://arxiv.org/abs/1008.1755}{\tt arXiv:1008.1755}] [hep-th].

\vspace{-2mm} 
\bibitem[Sa11c]{NS5}
H. Sati,  {\it Topological aspects of the NS5-brane},
[\href{https://arxiv.org/abs/0807.4940}{\tt arXiv:1109.4834}] [{\tt hep-th}].

\vspace{-2mm} 
\bibitem[Sa14]{sigma}	
H. Sati, M-theory with framed corners and tertiary index invariants, SIGMA 
 {\bf 10} (2014), 024, [\href{https://arxiv.org/abs/1203.4179}{\tt arXiv:1203.4179}].

 
 \vspace{-2mm} 
 \bibitem[SS18]{SS}
H. Sati and U. Schreiber,
{\it Higher T-duality of M-branes via local supersymmetry},
Phys. Lett. {\bf B 781} (2018), 694--698,
[\href{https://arxiv.org/abs/1805.00233}{\tt  arXiv:1805.00233}] [{\tt hep-th}].

 
  \vspace{-2mm} 
 \bibitem[SSS09]{SSS2}
 H. Sati, U. Schreiber, and J. Stasheff, {\it Fivebrane structures}, 
 Rev. Math. Phys. {\bf 21} (2009) 1-44, \newline
\href{https://arxiv.org/abs/0805.0564}{\tt	 arXiv:0805.0564}] [{\tt math.AT}].
 
 
  \vspace{-2mm} 
 \bibitem[SSS12]{SSS3}
H. Sati, U. Schreiber, and  J. Stasheff,
{\it Differential twisted String- and Fivebrane structures}, 
Commun.  Math. Phys. {\bf 315} (2012), 169--213,	
[\href{https://arxiv.org/abs/0910.4001}{\tt arXiv:0910.4001}] [{\tt math.AT}].


\vspace{-2mm} 
\bibitem[SWe15]{SWe}
H. Sati and C. Westerland,
{\it Twisted Morava K-theory and E-theory},
J. Topol. {\bf 8} (2015), no. 4, 887--916, \newline
[\href{https://arxiv.org/abs/1109.3867}{\tt  arXiv:1109.3867}] [{\tt math.AT}].


\vspace{-2mm} 
\bibitem[SW18]{SW1}
H. Sati and M. Wheeler, {\it Variations of higher rational tangential structures},
J. Geom. Phy. {\bf 130} (2018), 229-248,
%DOI:	10.1016/j.geomphys.2018.04.001,
[\href{https://arxiv.org/abs/1612.06983}{\tt	 arXiv:1612.06983}] [{\tt math.AT}].


\vspace{-2mm} 
\bibitem[SW19]{SW2}
H. Sati and M. Wheeler,
{\it Topological actions via gauge variations of higher structures}, 
Phys. Lett. {\bf B 789} (2019), 114-118,
[\href{https://arxiv.org/abs/1810.05349}{\tt		arXiv:1810.05349}] [{\tt hep-th}].

\vspace{-2mm} 
\bibitem[SW19b]{SW3}
H. Sati and M. Wheeler, {\it Higher structures and  6-dimensional $\mathcal{N}=(1,0)$ theory}, 
preprint. 

\vspace{-2mm} 
\bibitem[Se97]{Se}
N. Seiberg, {\it Nontrivial fixed points of the renormalization group in six-dimensions},
Phys. Lett. {\bf B 390} (1997) 169-171, 
[\href{https://arxiv.org/abs/hep-th/9609161}{\tt	arXiv:hep-th/9609161}].

\vspace{-2mm} 
\bibitem[SW96]{SWi}
 N. Seiberg and E. Witten, Comments on string dynamics in six-dimensions,
Nucl. Phys. {\bf B 471} (1996) 121-134,
[\href{https://arxiv.org/abs/hep-th/9603003}{\tt arXiv:hep-th/9603003}].


\vspace{-2mm} 
\bibitem[St96]{St}
 A. Strominger, {\it Open p-branes}, Phys. Lett. {\bf B383} (1996) 44-47,
[\href{https://arxiv.org/abs/hep-th/9512059}{\tt arXiv:hep-th/9512059}].


\vspace{-2mm} 
 \bibitem[Wa08]{Wa}
B.-L. Wang, {\it Geometric cycles, index theory and twisted 
K-homology},  J. Noncommut. Geom. {\bf 2} (2008), 497-552,
[\href{https://arxiv.org/abs/0710.1625}{\tt arXiv:0710.1625}][{\tt math.KT}]. 


\vspace{-2mm} 
 \bibitem[Wi95]{Wi}
E. Witten, {\it Some comments on string dynamics},
[\href{https://arxiv.org/abs/hep-th/9507121}{\tt  arXiv:hep-th/9507121}].


\end{thebibliography}
\end{document}